\documentclass[secnumeq,FBSmath]{FBSart}

\usepackage{graphicx}
\usepackage{bm}

\title{
Operation of Faddeev-Kernel in Configuration Space
}

\author{
S.\ Ishikawa\thanks{\textit{E-mail address}: ishikawa@hosei.ac.jp} 
}

\institute{
Department of Physics, Science Research Center, Hosei University, 
2-17-1 Fujimi, Chiyoda, Tokyo 102-8160, Japan
} 

\runningtitle{Faddeev-Kernel in Configuration Space}
\runningauthor{S.\ Ishikawa}

\begin{document}

\maketitle

\begin{abstract}
We present a practical method to solve Faddeev three-body equations at energies above three-body breakup threshold as integral equations in coordinate space. 
This is an extension of previously used method for bound states and scattering  states below three-body breakup threshold energy. 
We show that breakup components in three-body reactions produce long-range effects on Faddeev integral kernels in coordinate space, and propose numerical procedures to treat these effects. 
Using these techniques, we solve Faddeev equations for neutron-deuteron scattering to compare with benchmark solutions. 
\end{abstract}


%
%

\maketitle

\section{Introduction}

So far, a number of numerical methods to solve Faddeev three-body equations  
for energies above three-body breakup threshold have been developed and then applied to a system of nucleon-deuteron, which is considered as one of the most basic quantum three-body systems \cite{Fr90,Fr95}. 
These methods are classified into two groups: either to solve coupled integral equations for scattering amplitudes in momentum space or to solve coupled partial differential equations for wave functions in coordinate space. 

In this paper, we will present a different approach for three-body scattering problem above the breakup threshold energy, in which we solve the Faddeev integral equations for wave functions in coordinate space. 
This approach has been successfully applied to calculations of the three-nucleon bound states \cite{Sa86,Is86} and low-energy three-nucleon scattering below the breakup threshold energy with inclusion of three-nucleon forces \cite{Is86} and the long-range Coulomb interaction \cite{Sa81,Is03}. 

Integral equations for scattering problems are generally written in the form of  inhomogeneous linear equations.
In the previous works, we applied an iterative method called the method of continued fraction (MCF) to solve such equations, whose details are given in Refs. \cite{Sa86,Is87} and references therein. 
A basic procedure of the algorithm in the MCF is to operate the integral kernel to a function made in a preceding step, as are those in most iterative methods. 
It is thus essential to establish precise operations of integral kernels for solving the equations accurately, which is main subject of this paper.

The existence of three-body breakup channels causes some difficulties in three-body calculations. 
In the momentum space approach, for example, the effects appear as logarithmic singularities and discontinuities by a step function in the integral kernels of the equations so that we need to perform the integration very carefully \cite{Gl96}. 
In the differential equation approach, due to the breakup effects one needs to set boundary conditions at very long distance, the order of tens or hundreds times larger than the range of interaction potentials \cite{Ku87,Gl88,Gl92,Pa00}.
Since we treat the wave functions as solutions of the Faddeev integral equations, the long-range behavior should appear in the integral kernel.
In the present paper, we will describe how this behavior appears in our kernel, and how to treat it. 

Basic notations and steps of the kernel operation in detail are explained in Sec.\ \ref{sec:formulation} for a simple three-body system. 
In Sec.\ \ref{sec:numerical}, we show numerical examples of the kernel operation to a model function emphasizing some techniques to treat breakup effects, and then compare our calculations with benchmark tests \cite{Fr90,Fr95}.
Finally, we give a summary in Sec.\ \ref{sec:summary}.

\section{Formulation}
\label{sec:formulation}

\subsection{Notations}

Let us consider a system of three identical particles (nucleons) 1, 2, and 3.
We use sets of Jacobi coordinates \{$ \bm{x}_i, \bm{y}_i$\} defined as 
\begin{equation}
\left\{
\begin{array}{lll}
\bm{x}_i & = & \bm{r}_j - \bm{r}_k \\
\bm{y}_i & = & \bm{r}_i - \frac12 ( \bm{r}_j + \bm{r}_k )
\end{array}
\right.,
\end{equation}
where $(i,j,k)$ denotes (1,2,3) or its cyclic permutations and $\bm{r}_i$ is the position vector of the nucleon $i$.
For simplicity, we assume that the nucleons $j$ and $k$ interact via a short range pair wise potential $V_i = V(x_i)$, where $x_i = \vert \bm{x}_i\vert$, and that the potential supports a $s$-wave bound state (the deuteron) of energy $E_d$, whose radial part wave function is denoted by $\phi^d(x_i)$. 

We are going to obtain a wave function $\Psi$ corresponding to a scattering process initiated by a state with a nucleon and a deuteron having relative momentum $p_0$.
Faddeev equations for the process in the form of integral equations are 
\begin{equation}
 \Phi_i  =  \Xi_i + G_i V_i \left( \Phi_j + \Phi_k \right) = 
 \Xi_i + G_i V_i \hat{P} \Phi_i,
\label{eq:Fad}
\end{equation}
where $\Phi_i$'s are Faddeev components to make $\Psi$ as
\begin{equation}
\Psi = \Phi_1 + \Phi_2 + \Phi_3, 
\end{equation}
$\Xi_i$ is an initial state consisting of the deuteron for the pair $(j,k)$ and incoming free nucleon $i$, 
and
$G_i$ is a three-body channel Green's operator with the outgoing boundary condition,
\begin{equation}
G_i = \frac1{E + \imath \varepsilon 
 + \frac{\hbar^2}{m} \nabla^2_{x_i} +\frac{3\hbar^2}{4m} \nabla^2_{y_i} - V_i}. 
\end{equation}
The total energy in the three-body center of mass (c.m.) frame $E$ is given as 
\begin{equation}
E =  \frac{3 \hbar^2}{4m} p_0^2 + E_d 
 = \frac{3 \hbar^2}{4m} p_0^2 - \vert E_d \vert, 
\end{equation}
where $m$ denotes the nucleon mass. 
The operator $\hat{P}$ represents permutations of the particle numbers,
\begin{equation}
\hat{P} \Phi_i = \Phi_j  + \Phi_k. 
\end{equation}
A partial wave decomposition is performed by introducing an angular function ${\cal Y}_\alpha(\hat{\bm{x}}_i, \hat{\bm{y}}_i)$, 
\begin{equation}
  {\cal Y}_\alpha(\hat{\bm{x}}_i, \hat{\bm{y}}_i)   = 
\left[ Y_L(\hat{\bm{x}}_i) \otimes Y_\ell(\hat{\bm{y}}_i) \right]_{M_0}^{J_0},
\end{equation}
where $L$ denotes the relative orbital angular momentum of the pair $(j,k)$; 
$\ell$ the orbital angular momentum of the spectator $i$ with respect to the c.m. of the pair $(j,k)$; 
$J_0$ the total angular momentum of the three-body system ($\bm{J}_0 = \bm{L}+\bm{\ell}$); 
$M_0$ the third component of $J_0$.
The set of the quantum numbers $(L, \ell, J_0, M_0)$ are represented by the index $\alpha$. 
Furthermore, we use an index $\alpha_0$ to denote an initial partial wave state specifically with $L=0$.

\subsection{Kernel Operation}
\label{sec:kernel}

In this subsection, we describe how to handle the operation of the Faddeev kernel $G V \hat{P}$ on a given function $\Xi$,
\begin{equation}
\langle \bm{x}, \bm{y} \vert \Xi \rangle = 
 \sum_{\alpha} {\cal Y}_{\alpha}(\hat{\bm{x}}, \hat{\bm{y}}) \xi_\alpha (x,y)
\label{eq:Xi-pwd}
\end{equation}
to produce a new function $\Phi$,
\begin{eqnarray}
\langle \bm{x}, \bm{y} \vert \Phi \rangle 
 &=& \langle \bm{x}, \bm{y} \vert  GV \hat{P} \vert \Xi \rangle
\nonumber \\
&=& 
 \sum_{\alpha} {\cal Y}_{\alpha}(\hat{\bm{x}}, \hat{\bm{y}}) \phi_\alpha (x,y),
\label{eq:GVpi}
\end{eqnarray}
where we have dropped the particle number indices $(i,j,k)$ for simplicity.

The kernel operation starts with the permutation operator $\hat{P}$ to define a function $\chi_\alpha(x,y)$,
\begin{equation}
\chi_\alpha(x,y) = ( {\cal Y}_{\alpha} \vert \hat{P} \vert \Xi \rangle.
\label{eq:chi_xy}
\end{equation}
In the case of identical particles, $\hat{P}$ is nothing but a coordinate exchange operator, whose operations are summarized in \ref{sec:OperatorP}.

Next step is the operation of the Green's operator $G$.
In the case of the scattering problem, where $E>0$, the Green's operator $G$ possesses a pole corresponding to the deuteron bound state. 
In order to treat this pole, we apply a standard subtraction method, in which we insert a trivial identity, 
\begin{equation} 
1 = \sum_{\alpha_0} \vert{\cal Y}_{\alpha_0} \phi^d ) (\phi^d {\cal Y}_{\alpha_0}\vert 
  +  \left[ 1 -  \sum_{\alpha_0} \vert{\cal Y}_{\alpha_0}\phi^d) (\phi^d{\cal Y}_{\alpha_0}\vert \right], 
\end{equation} 
between $G$ and  $V$ in Eq.\ (\ref{eq:GVpi}).
This procedure extracts an elastic contribution of the Green's operator \cite{Sa78} and leads to an expression, 
\begin{equation}
\phi_\alpha(x,y) = \delta_{\alpha,\alpha_0}\phi^d(x) {\cal F}^{(e)} (y)
   + \phi_\alpha^{(b,c)}(x,y).
\label{eq:phi_x_y}
\end{equation}

Here, ${\cal F}^{(e)}(y)$ represents an elastic component in the scattering given by
\begin{equation}
{\cal F}^{(e)}(y)
 =  \int_0^\infty y^{\prime 2} dy^\prime 
  \breve{G}_{0,\ell_0}(y, y^\prime) \omega^{(e)}(y^\prime),
\label{phi-e}
\end{equation}
where $\breve{G}_{0,\ell}(y, y^\prime)$ is a partial wave component of the free Green's operator for the outgoing particle, 
\begin{eqnarray}
 \breve{G}_{0,\ell}(y, y^\prime ) &\equiv&
 \left( y \left\vert \frac1{\frac{3\hbar^2}{4m} p_0^2 + \imath \varepsilon - T_\ell(y)} \right\vert y^\prime \right)
\nonumber\\
&=&
-\frac{4m}{3\hbar^2} p_0 
  h^{(+)}_\ell(p_0y_{>}) j_\ell(p_0y_{<}) 
\label{eq:breveG}
\end{eqnarray}
with
\begin{equation}
T_\ell(y) = -\frac{3 \hbar^2}{4 m} \left(
   \frac{d^2}{dy^2} + \frac2y \frac{d}{dy} - \frac{\ell(\ell+1)}{y^2} \right).
\end{equation}
In Eq. (\ref{eq:breveG}), $j_\ell(p_0 y)$ is the spherical Bessel function and 
$h^{(+)}_\ell(p_0 y)$ is the spherical Hankel function with the outgoing wave, where the outgoing $(+)$ and the incoming $(-)$ spherical Hankel functions are defined with the spherical Neumann function $n_\ell(p_0y)$ as 
\begin{equation}
h^{(\pm)}_\ell(x) = -n_\ell(x) \pm \imath j_\ell(x).
\end{equation}

The function $\omega^{(e)}(y)$, which plays a role of the source for the elastic component in Eq. (\ref{phi-e}), is given by
\begin{equation}
 \omega^{(e)}(y)
=  \int_0^\infty x^2 dx \phi^d(x) V(x) \chi_{\alpha_0}(x,y).
\label{eq:omega_y}
\end{equation}

The explicit expression of the Green's function Eq. (\ref{eq:breveG}) gives the asymptotic form of ${\cal F}^{(e)}(y)$  as
\begin{equation}
{\cal F}^{(e)}(y) \mathop{\to}_{y \to \infty} h_{\ell_0}^{(+)}(p_0 y)~ T^{(e)},
\label{eq:phi_e_asym}
\end{equation}
where $T^{(e)}$ is the elastic $T$-matrix amplitude defined by
\begin{equation}
T^{(e)} = - p_0 \left( \frac{4m}{3 \hbar^2}\right) 
    \int_0^\infty y^2 dy j_{\ell_0}(p_0 y)  \omega^{(e)}(y).
\end{equation}

The second term in the right hand side of Eq.\ (\ref{eq:phi_x_y}) expresses three-body breakup and closed-channel components in the scattering. 
In our formalism, these components are treated by expanding the Faddeev kernel with respect to a spectator particle state of momentum $p$, 
\begin{equation}
u_\ell (y;p) \equiv \sqrt{\frac2{\pi}} p j_\ell (py), 
\end{equation}
which satisfies a complete relation
\begin{equation}
\frac{\delta(y-y^\prime)}{y y^\prime}
 = \int_0^\infty dp u_\ell (y;p) u_\ell (y^\prime;p).
\end{equation}
The function $\phi_{\alpha}^{(b,c)}(x,y)$  thereby is written as a Fourier-Bessel transformation:
\begin{equation}
\phi_{\alpha}^{(b,c)}(x,y) 
= \int_0^{\infty} dp u_{\ell}(y;p) 
  \Big[  \eta_{\alpha}(x;p) 
  -\delta_{\alpha, \alpha_0} \phi^d(x) C_{\alpha}(p) \Big].
\label{eq:phi-bc}
\end{equation}
Here, $\eta_{\alpha}(x;p)$ is defined as
\begin{equation}
\eta_{\alpha}(x;p)
 = \langle x \vert G_{L} \vert \hat{\omega}_{\alpha} \rangle,
\label{eq:eta-Gomega}
\end{equation}
where $G_{L}$ is a two-body Green's operator 
\begin{equation}
G_{L} =  \frac1{E_q + \imath \varepsilon - T_{L}(x) - V(x)} 
\label{eq:Green}
\end{equation}
with
\begin{equation}
T_L(x) = -\frac{\hbar^2}{m} \left(
   \frac{d^2}{dx^2} + \frac2x \frac{d}{dx} - \frac{L(L+1)}{x^2} \right).
\end{equation}
The energy of the two-body subsystem $E_q$ is given by 
\begin{equation}
E_q = E - \frac{3 \hbar^2}{4m} p^2 = \frac{\hbar^2}{m} q^2, 
\label{eq:Eq}
\end{equation}
and the $p$-dependence of the functions arises through this relation.

The breakup component stems from the integral of the first term in Eq.\ (\ref{eq:phi-bc}) for the range of $0 \le p \le p_c = \sqrt{4mE/3\hbar^2}$. 
In this range the energies of both the spectator particle and the two-body subsystem are positive or zero, and thus the integral survives at infinite values of $x$ and $y$, see Eq.\ (\ref{eq:phi_asym}) below and Refs. \cite{Sa77,Ru65}. 
The rest of the integral of the first term in Eq.\ (\ref{eq:phi-bc}), i.e., $p_c < p < \infty$, as well as the second term in Eq.\ (\ref{eq:phi-bc}) damp for large values of $x$ and $y$ because the energy of the two-body subsystem is negative. 
In this sense, we call these components {\em closed}.

The source term in Eq. (\ref{eq:eta-Gomega}), $\hat{\omega}_{\alpha}(x;p)$, is written as
\begin{equation}
\hat{\omega}_{\alpha}(x;p) = V(x) \hat{\chi}_{\alpha}(x;p),
\label{eq:omega_xp}
\end{equation}
\begin{equation}
\hat{\chi}_{\alpha}(x;p) 
 = \int_0^\infty y^2 dy u_{\ell}(y;p) \chi_{\alpha}(x,y).
\label{eq:chi_xp}
\end{equation}

The second term of the right hand side in Eq.\ (\ref{eq:phi-bc}) appears as a counter part of the subtraction and $C_{\alpha}(p)$ is defined as
\begin{equation}
C_{\alpha}(p)  =  \frac1{E_{q} - E_d}
\int_0^\infty x^2 dx \phi^d(x) \hat{\omega}_{\alpha}(x;p). 
\end{equation}
The apparent singularity in $C_{\alpha}(p)$ cancels that of the two-body Green's operator $G_{L}$, which will be numerically shown in the following section, and thus, we can apply a standard quadrature to perform the $p$-integration 
in Eq.\ (\ref{eq:phi-bc}) as far as the both terms are treated together.

In calculating $\eta_{\alpha}(x;p)$, we transform Eq.\ (\ref{eq:eta-Gomega}) to an ordinary differential equation:
\begin{equation}
\left[ E_q  - T_{L}(x) - V(x) \right] \eta_{\alpha}(x;p) 
 = \hat{\omega}_{\alpha}(x;p)
\label{eq:SchFad}
\end{equation}
with boundary conditions
\begin{equation}
\eta_{\alpha}(x;p) 
   \mathop{\propto}_{x \to \infty}
\left\{
\begin{array}{ll}
  h^{(+)}_{L} ( q x) & (0 \le p \le p_c)
\\
  h^{(+)}_{L} ( \imath \vert q \vert x)& ( p_c < p < \infty)
\end{array}
\right..
\label{eq:bd-cond}
\end{equation}
A treatment of the two-body Green's operator at three-body breakup region, $0 \le p \le p_c$ will be described in \ref{sec:Green-x}.
We here only note that the asymptotic form of $\eta_{\alpha}(x;p)$ is given by
\begin{equation}
\eta_{\alpha}(x;p)  \mathop{\to}_{x \to \infty} 
 h_{L}^{(+)}(qx) \frac{\left( -q \frac{m}{\hbar^2} \right)}{1- \imath {\cal K}_{L}(q)}
 \langle \hat{\psi}_{L}(q) \vert \hat{\omega}_{\alpha} \rangle,
\label{eq:eta_asym}
\end{equation}
where $\hat{\psi}_{L}(x;q)$ is a two-body scattering solution with the standing wave boundary condition and ${\cal K}_{L}(q)$ is a scattering $K$-matrix for the two-body scattering (See \ref{sec:Green-x}).


The asymptotic form of $\phi_{\alpha}^{(b,c)}(x,y)$ is evaluated by the saddle-point approximation \cite{Sa77,Ru65} as 
\begin{equation}
\phi_{\alpha}^{(b,c)}(x,y) \mathop{\to}_{x \to \infty, x/y~\mathrm{fixed}} 
- e^{\frac{\pi}{4} \imath} \imath^{-L-\ell} \left(\frac{4K_0}{3}\right)^{3/2}
  \frac{e^{\imath K_0 R}}{R^{5/2}}
  B_{\alpha}(\Theta),
\label{eq:phi_asym}
\end{equation}
where we introduce a hyper radius $R$ and a hyper angle $\Theta$ as
\begin{equation}
R = \sqrt{x^2 + \frac43 y^2},
\end{equation}
\begin{equation}
x = R \cos\Theta,~~~~y= \sqrt{\frac34} R \sin\Theta,
\end{equation}
and $K_0$ is given by
\begin{equation}
K_0 = \sqrt{\frac{m}{\hbar^2} E}.
\end{equation}

$B_{\alpha}(\Theta)$ is the breakup amplitude defined as
\begin{equation}
B_{\alpha}(\Theta) = 
  - \frac1{\bar{p}} \frac{m}{\hbar^2} \frac{1}{1-\imath {\cal K}_{L}(\bar{q})}
  \langle \hat{\psi}_{L}(\bar{q}) \vert
   \hat{\omega}_{\alpha} \rangle.
\label{def:B_amp}
\end{equation}
Here, the momenta $\bar{q}$ and $\bar{p}$ are given as
\begin{equation}
 \bar{q} = K_0 \cos\Theta, \qquad \bar{p}= \sqrt{\frac43} K_0 \sin\Theta.
\end{equation}

\section{Numerical Analyses and Results}
\label{sec:numerical}

\subsection{Model source function}

In this section, we present a numerical example of the kernel operation described in the preceding section with a model source function that is restricted to $L=\ell=J_0=0$ state but carries a feature of the presence of three-body breakup channel similarly as the one used in Ref.\ \cite{Pa00}:
\begin{equation}
\chi_\alpha(x,y) =  \frac{e^{i K_0 R}}{(R+R_0)^{5/2}} 
\label{eq:chi_def}
\end{equation}
with $R_0= 5$ fm.

We choose the ${}^3$S$_1$-component of the Malfliet-Tjon model as presented in Ref.\ \cite{Fr90} for the potential $V(x)$ and the incident nucleon energy of $E_{Lab}=14.1$ MeV, which gives $K_0$=0.416 fm$^{-1}$, $p_c$=0.480 fm$^{-1}$, and $p_0$=0.550 fm$^{-1}$.
In numerical calculations below, mesh points for $x$- and $y$-variables in described in \ref{sec:meshes} are used.

\subsection{Elastic part}

\begin{figure}[tbh]
\includegraphics[scale=1]{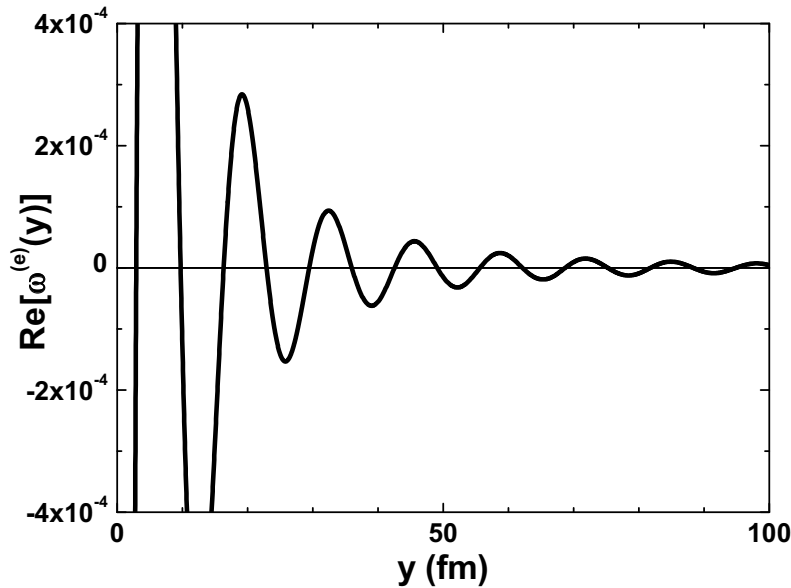}
\caption{
Real part of the elastic source function $\omega^{(e)}(y)$.
\label{fig:omegay}}
\end{figure}

In Fig.\ \ref{fig:omegay}, we plot the real part of the elastic source function $\omega^{(e)}(y)$,  Eq. (\ref{eq:omega_y}), calculated with the model function 
Eq.\ (\ref{eq:chi_def}). 
As shown in this figure, $\omega^{(e)}(y)$ reveals a  long-range behavior, which is given by
\begin{equation}
\omega^{(e)}(y) \mathop{\propto}_{y \to \infty} 
\frac{e^{\imath p_c y}}{y^{5/2}},
\label{eq:omega_aymp}
\end{equation}
whose oscillation length $\frac{2\pi}{p_c}$ is about 13 fm.

In calculating the elastic component ${\cal F}^{(e)}(y)$,
we treat this long-range behavior by rewriting  Eq. (\ref{phi-e}) as
\begin{equation}
{\cal F}^{(e)}(y) = - n_{\ell_0}(p_0 y) T(y)
 + \imath j_{\ell_0}(p_0 y) T^{(e)}
 + j_{\ell_0}(p_0 y) \left(S(y) - \hat{S} \right),
\label{eq:F_elST}
\end{equation}
where we have defined $T(y)$,  $S(y)$, and $\hat{S}$ by 
\begin{eqnarray}
T(y) &=& - p_0 \left(\frac{4m}{3\hbar^2}\right) 
       \int_0^y y^{\prime 2} dy^\prime j_{\ell_0}(p_0 y^\prime) \omega^{(e)}(y^\prime)
\mathop{\to}_{y \to \infty} T^{(e)}
\nonumber \\
S(y) \ &=& - p_0 \left(\frac{4m}{3\hbar^2}\right) 
  \int_0^y y^{\prime 2} dy^\prime n_{\ell_0}(p_0y^\prime) \omega^{(e)}(y^\prime)
\mathop{\to}_{y \to \infty} \hat{S}.
\label{def:TySy}
\end{eqnarray}

In numerical integration in Eq.\ (\ref{def:TySy}), we need to be careful for oscillational behaviors of the spherical Bessel and spherical Neumann functions as well as $\omega^{(e)}(y)$. 
This is done by spline interpolation technique used in Ref.\ \cite{Sa81} taking into account of oscillational behavior of the integrand carefully.

\begin{figure}[tbh]
\includegraphics[scale=1]{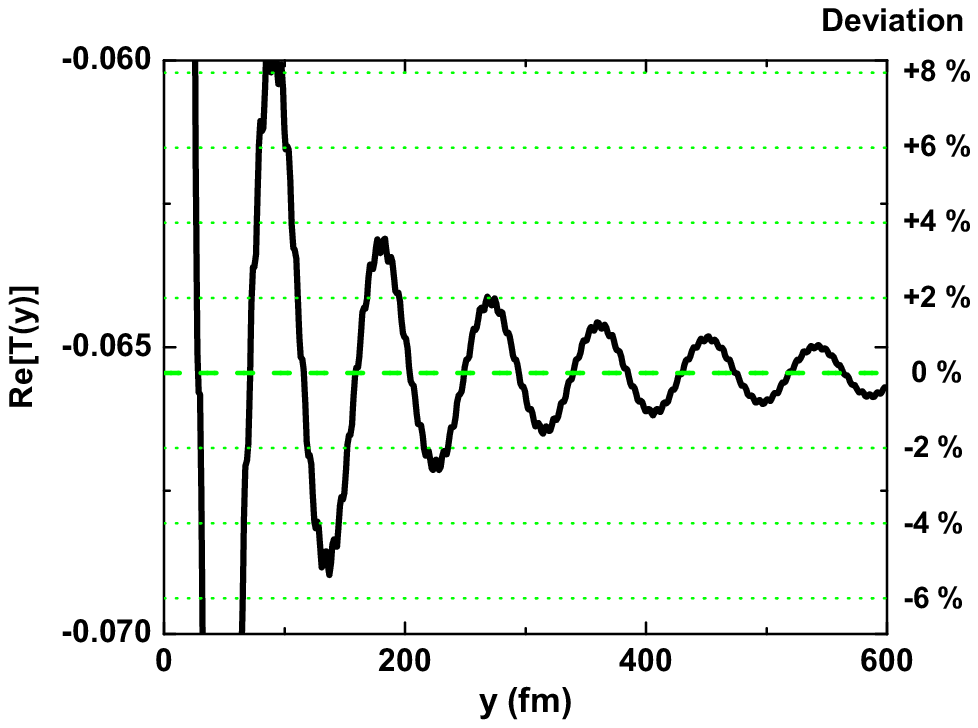}
\caption{
The real part of the function $T(y)$.
The obtained converged value is shown by the dashed line.
Dotted lines with the indices in the right hand side axis denote the deviation of the real part of $T(y)$ from the converged value.
\label{fig:T_y}}
\end{figure}

For the use of Eq. (\ref{eq:F_elST}), one needs converged values of $T(y)$ and $S(y)$ for $y\to\infty$. 
In Fig.\ \ref{fig:T_y}, we plot the real part of $T(y)$ for an example. 
As is expected from the long-range behavior of $\omega^{(e)}(y)$, the convergence of $T(y)$ becomes very slow. 
However, from the functional form of Eq.\ (\ref{eq:omega_aymp}), we expect that the function $T(y)$ behaves asymptotically as
\begin{equation}
T(y) \mathop{\to}_{y \to \infty}  t_0 + t_1 \frac{e^{\imath (p_0-p_c)y}} {y^{3/2}},
\label{eq:T_asym}
\end{equation}
where $t_0$ and $t_1$ are expansion coefficients and the coefficient $t_0$ is considered as a converged value of $T^{(e)}$.
The wave length evaluated by this equation is $\frac{2\pi}{p_0-p_c}=90$ fm, which is actually observed in Fig. \ref{fig:T_y}.
The fitting coefficients in Eq.\ (\ref{eq:T_asym}) are evaluated by a least square fit. 
To do this, the calculated values of $T(y)$ in a range of 80 $\le y \le y_{fit}^{max}$ (fm) are used. 
In Table \ref{tab:Ty_conv}, the dependence of the result on $y_{fit}^{max}$ is displayed.
From the table, we set $y_{fit}^{max}=1000$ fm to get a converged result in five digits of accuracy, which is denoted by the dashed line in Fig.\ \ref{fig:T_y}.
We remark that this result is obtained in spite of the fact the deviation of $T(y)$ from the converged value is still about  0.5 \% at  $y=1000$ fm, which is not shown in Fig. \ref{fig:T_y}.

\begin{table}
\beforetab
\begin{tabular}{cl}
\firsthline
$y_{fit}^{max} $ (fm) & Re$[t_0]$   \\
\midhline
532 &  -6.5454  \\
622 &  -6.5452  \\
983 &  -6.5449  \\
1073 & -6.5448 \\
1163 & -6.5448 \\
1193 & -6.5448 \\
\lasthline
\end{tabular}
\aftertab
\captionaftertab[]{
The real part of the fitting coefficient $t_0$.
\label{tab:Ty_conv}}

\end{table}

Here, we consider a range of the variables $\{x,y\}$ to be used in calculations. 
In the Faddeev equation, the function $\chi_\alpha(x,y)$ is always accompanied by the potential $V(x)$, which means that we need to calculate this function within the range of potential, $x_R$, for the $x$-variable. 
On the other hand, there is no restriction for the $y$-variable. 
In actual, Table \ref{tab:Ty_conv} demonstrates that we need to calculate $\chi_\alpha(x,y)$ for a large value of $y$, i.e., 1000 fm. 

Suppose that we calculate $\chi_\alpha(x,y)$ by Eqs.\ (\ref{eq:Xi-pwd}) and (\ref{eq:chi_xy}) with a function $\xi_\alpha(x,y)$, which is obtained in a preceding iteration step, for a range of $\left\{ 0 \le x \le x_R, 0 \le y \le y_{fit}^{max} \right\}$. 
The formulae, Eqs.\ (\ref{eq:K_gamma}) and (\ref{eq:xy-prime}), show that we need to prepare the function $\xi_\alpha(x,y)$ for a range of 
\begin{eqnarray}
0 \le &x& \le \frac12 x_R + y_{fit}^{max}
\nonumber\\
0 \le &y& \le \frac34 x_R + \frac12 y_{fit}^{max}
\label{eq:xy-range}
\end{eqnarray}
to perform the exchange operation to obtain $\chi_\alpha(x,y)$ for the above range. 
If we set $x_R$=10 fm  and $y_{fit}^{max}$=1000 fm, this turns to be 
$\{ 0 \le x \le 1005~(\mbox{fm}), 0 \le y \le 507.5~(\mbox{fm}) \}$, 
which is rather huge.

To facilitate numerical calculations, we limit the range of calculating  $\chi_\alpha(x,y)$ to $\{ 0 \le x \le x_R, 0 \le y \le y_M \}$ by choosing the value of $y_M$ adequately, and approximate the value of $\chi_\alpha(x,y)$ for $y > y_M$ using a form of 
\begin{equation}
\chi_\alpha(x,y) \mathop{=}_{0 \le x \le x_R, y \ge y_M} 
  \frac{e^{\imath \sqrt{\frac43} K_0 y}}{y^{2/5}} 
\left( a_0(x) + \frac{a_1(x)}{y} + \frac{a_2(x)}{y^2}\right),
\label{eq:chi_asm}
\end{equation}
where the coefficients $a_n(x)$ are determined by a least square fit to  $\chi_\alpha(x,y)$ for $y<y_M$ and for each value of $x$.

With a choice of $x_M=10$ fm and $y_M=80$ fm, by which the range for $\xi_\alpha(x,y)$ becomes $\{0 \le x \le 85~\mbox{(fm)}, 0 \le y \le 47.5~\mbox{(fm)} \}$, we obtain the equivalent results for $T(y)$ and its asymptotic value $T^{(e)}$ to the previously shown. 
This procedure reduces the amount of calculations considerably without loss of accuracy, and will be applied in the following analyses. 

\begin{figure}[tbh]
\includegraphics[scale=1]{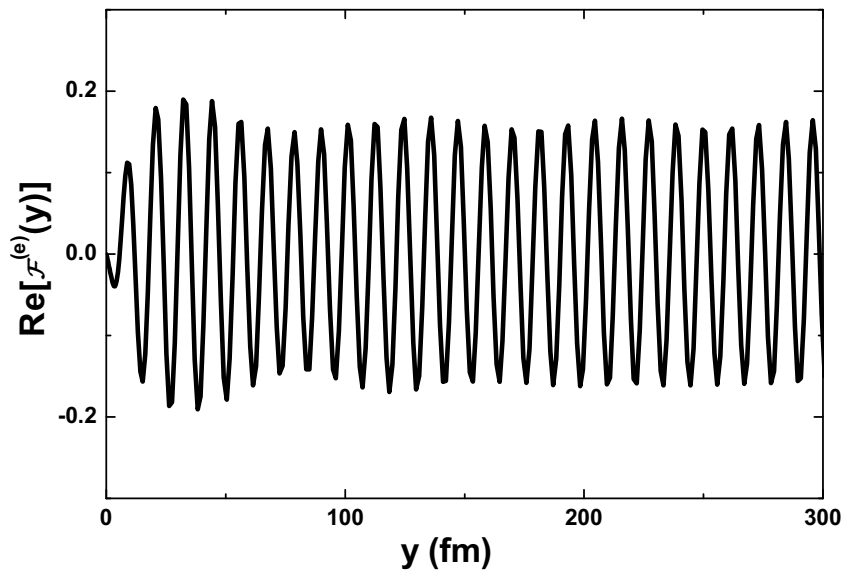}
\caption{
The real part of the elastic function ${\cal F}^{(e)}(y)$.
\label{fig:elasticf}}
\end{figure}

Together with the function $S(y)$ and its asymptotic value $\hat{S}$ calculated similarly, the elastic component ${\cal F}^{(e)}(y)$ is constructed using Eq.\ (\ref{eq:F_elST}), whose real part is plotted in Fig. \ref{fig:elasticf}. 
Note that the effect of the slow convergence in $T(y)$ and $S(y)$ functions appears as a small oscillation of the amplitude of ${\cal F}^{(e)}(y)$ with the wave length of $\frac{2\pi}{p_0-p_c}=90$ fm, which exists up to a large distance where the convergences of $T(y)$ and $S(y)$ are achieved. 
%

\subsection{Breakup and closed channel parts}

First step in the calculation of the three-body breakup and closed channel contributions is the Fourier-Bessel transformation of $\chi_\alpha(x,y)$ with respect to the coordinate $y$, Eq.\ (\ref{eq:chi_xp}). 
We again face the problem of slow convergence in the $y$-integral due to the long-rangeness of $\chi_\alpha(x,y)$.
This is treated similarly with the calculation of the elastic component by writing Eq.\ (\ref{eq:chi_xp}) as 
\begin{eqnarray}
\hat{\chi}_\alpha(x;p) 
  &=& \int_0^{y_M} y^{\prime 2}dy^\prime u_\ell(y^\prime;p)  \chi_\alpha(x,y^\prime)
\nonumber\\
 && + \lim_{y\to\infty} 
      \int_{y_M}^{y} y^{\prime 2}dy^\prime u_\ell(y^\prime;p)  \chi_\alpha(x,y^\prime).
\label{eq:chi_div}
\end{eqnarray}

\begin{figure}[tbh]
\includegraphics[scale=1]{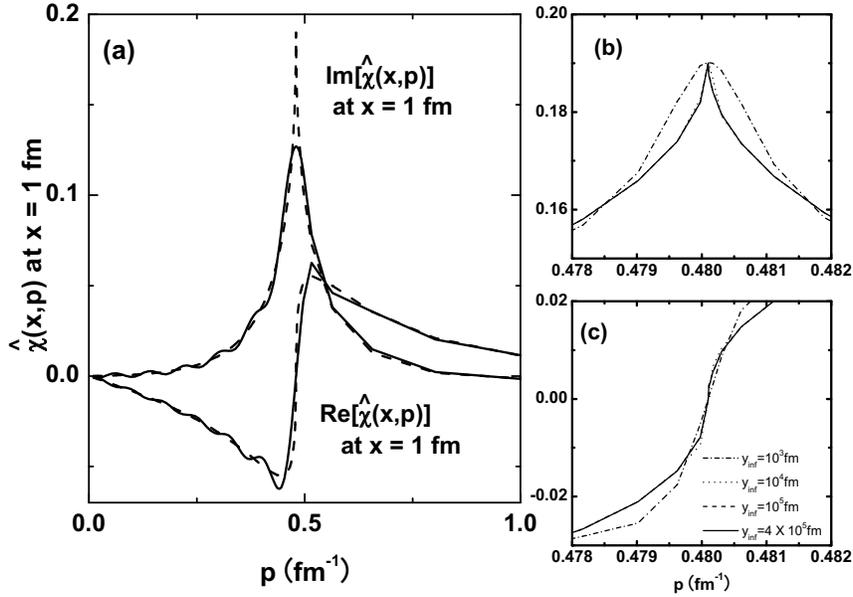}
\caption{
(a) The real and imaginary parts of $\hat{\chi}_\alpha(x;p)$ for $0 \le p \le 1$ (fm${}^{-1}$) at $x=$ 1 fm.
The solid curves are the first term of Eq. (\protect\ref{eq:chi_div}) with $y_M=80$ fm.
The dashed curves are the full calculation.
(b) The imaginary part and (c) the real part of  $\hat{\chi}_\alpha(x;p)$ around $p=p_c$ at $x=$ 1 fm for various values of $y_{inf}$.
See the text for the details.
\label{fig:chixpYM}}
\end{figure}

The first term is integrated numerically using the spline interpolation technique \cite{Sa81}. 
Results for the real and the imaginary parts of the integrals with $y_M=80$ fm are shown in Fig. \ref{fig:chixpYM} (a) by the solid curves. 
The oscillational behavior of the curves indicates that the integrals do not converge yet.

In calculating the second term in Eq.\ (\ref{eq:chi_div}), we use the asymptotic form of $\chi_\alpha(x,y)$ given by Eq.\ (\ref{eq:chi_asm}).
Now, we define a function ${\cal I}^{(n)}(y;p)$ ($n=$0, 1, or 2) as
\begin{equation}
{\cal I}^{(n)}(y;p) = \int_{y_M}^{y} dy^\prime  y^\prime u_\ell(y^\prime;p) 
 \frac{e^{\imath p_c y^\prime}} {(y^\prime)^{3/2+n}}, 
\label{eq:chi_asyp_int}
\end{equation}
and then express this for large values of $y$ in a form of 
\begin{equation}
b_0^{(n)}(p) 
  + b_1^{(n)}(p) \frac{e^{\imath (p-p_c) y}}{y^{3/2+n}}  
\label{eq:chi_asym}
\end{equation}
with expansion coefficients $b_0^{(n)}$ and $b_1^{(n)}$ to be determined by a least square fit. 
The wave length of the oscillation of Eq.\ (\ref{eq:chi_asym}) with respect to $y$-variable depends on the momentum $p$ as $\frac{2\pi}{p-p_c}$.
In a particular case of $p=p_c$, where no oscillation occurs, a functional form of
\begin{equation}
b_0^{(n)} + \frac{b_1^{\prime (n)}}{y^{1/2+n}} 
  + \frac{b_2^{\prime (n)}}{y^{3/2+n}}
\label{eq:chi_asym0}
\end{equation}
is used.
The second term of Eq.\ (\ref{eq:chi_div}) is thereby expressed as
\begin{equation}
  \sum_{n=0}^{2} a_n(x) b_0^{(n)}(p).
\label{eq:int_fit}
\end{equation}

%
Since the functions  ${\cal I}^{(n)}(y;p)$ depend only on $y_M$ and the total energy $E$, we may calculate them once in advance to start an iterative process in solving the Faddeev equations.

The $b$-coefficients in Eq.\ (\ref{eq:chi_asym}) are obtained from a least square fit using values of ${\cal I}^{(n)}(y;p)$ for a range up to $y=y_{inf}$. 
To obtain accurate values of the coefficients, we need to include at least several oscillations in the range. 
Since the wave length of the oscillation becomes larger as $p$ approaching to $p_c$, the maximum value $y_{inf}$ to get a converged result could become a huge number. 
This is illustrated in Figs. \ref{fig:chixpYM} (b) and (c), where the dependence of the resultant $\hat{\chi}_\alpha(x;p)$ on some selected values of $y_{inf}$ is plotted.
In the figures, we plot the real and imaginary parts of $\hat{\chi}_\alpha(x;p)$ around $p=p_c=0.480$ fm${}^{-1}$ at $x=1$ fm calculated by choosing $y_{inf}$ = $10^3$ fm (dot-dashed curves), $10^4$ fm (dotted curves), $10^5$ fm (dashed curves), and $4\times10^5$ fm (solid curves).
One sees that even the value of $y_{inf}=10^3$ fm is not enough to get a converged result.
Numerically, it turns out that $4\times10^5$ fm may be good enough.
The results with $y_{inf}=4\times10^5$ fm are plotted as dotted curves in Fig. \ref{fig:chixpYM} (a).
The oscillating behavior due to the small value of the integral maximum given by the solid curves disappears by taking into account of the long-range character of the source function $\chi_\alpha(x,y)$.

Using thus obtained $\hat{\chi}_\alpha(x;p)$, one calculates $\hat{\omega}_{\alpha}(x;p)$ from Eq.\ (\ref{eq:omega_xp}), and then solves the ordinary differential equation, Eq.\ (\ref{eq:SchFad}), with the boundary conditions Eq.\ (\ref{eq:bd-cond}) to obtain $\eta_\alpha(x;p)$. 
The Numerov algorithm is applied for solving this equation as in Refs.\ \cite{Sa79,Sa81} with $x$-mesh points described in \ref{sec:meshes}. 

\begin{figure}[htb]
\includegraphics[scale=1]{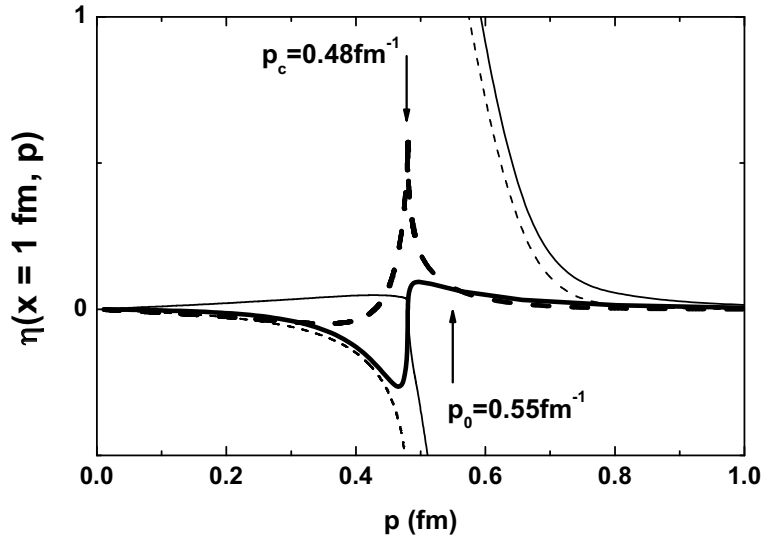}
\caption{
Thin curves are the real part (solid curve) and the imaginary part (dashed curve) of $\eta_\alpha(x;p)$ at $x=1$ fm as functions of $p$.
Bold curves are those after the subtraction of $\phi^d(x) C_\alpha(p)$-term in Eq.\ (\protect\ref{eq:phi-bc}).
\label{fig:fpixp}
}
\end{figure}

Fig.\ \ref{fig:fpixp} displays the real (imaginary) part of the resultant $\eta_\alpha(x;p)$ function at $x=1$ fm as thin solid (thin dashed) curve.
The discontinuous singularities of thin curves at $p=p_0=0.550$ fm${}^{-1}$ correspond to the deuteron pole in the two-body Green's function. 
These singularities disappear when the term of $\phi^d(x) C_\alpha(p)$ in Eq.\ (\ref{eq:phi-bc}) is subtracted, as shown as bold curves in the figure.

\begin{figure}[tbh]
\includegraphics[scale=1]{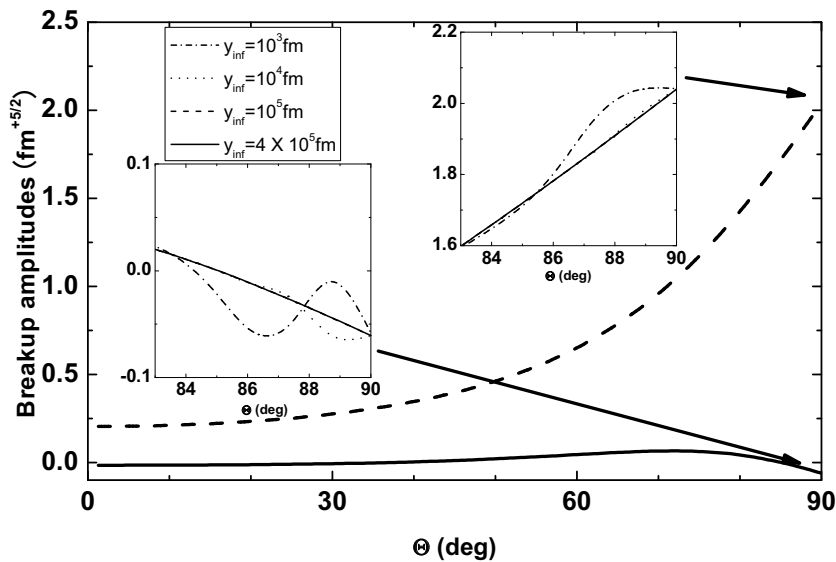}
\caption{
The breakup amplitude $B_\alpha(\Theta)$. 
The solid line shows the real part and the dashed line the imaginary part.
The inserts show the behavior of $y_{inf}$-dependence in calculating $\chi_\alpha(x;p)$ function.
The meaning of each curve is explained in the text.
\label{fig:bampym}
}
\end{figure}

In our formalism, the breakup amplitude $B_\alpha(\Theta)$ is obtained by two different ways. 
One way is to use Eq.\ (\ref{def:B_amp}) directory, which can be performed before solving the ordinary differential equation Eq. (\ref{eq:SchFad}). 
After getting a solution of Eq. (\ref{eq:SchFad}), the breakup amplitude is calculated from its asymptotic form Eq.\ (\ref{eq:eta_asym}) as the second way.
Both calculations agree each others, which assures the accuracy of the solutions of Eq. (\ref{eq:SchFad}), and displayed in Fig. \ref{fig:bampym}. 
In the inserts of Fig. \ref{fig:bampym}, results with different values of $y_{inf}$ in calculating $\hat{\chi}_\alpha(x;p)$ are displayed as in Fig. \ref{fig:chixpYM} to see effects of the long-range properties of $\chi_\alpha(x,y)$ in the region of $\Theta \sim \pi/2$ region, where $q \sim 0$. 

Once the function $\eta_\alpha(x;p)$ is obtained, by performing the transformation Eq.\ (\ref{eq:phi-bc}) with the spline interpolation technique, we obtain the function $\phi_\alpha^{(b,c)}(x,y)$.
Together with the elastic component ${\cal F}^{(e)}(y)$, we finally obtain $\phi(x,y)$ by Eq.\ (\ref{eq:phi_x_y}).

\subsection{Comparison with the Benchmark solutions}

\begin{figure}[tbh]

\begin{minipage}[t]{65mm}
\includegraphics[scale=0.8]{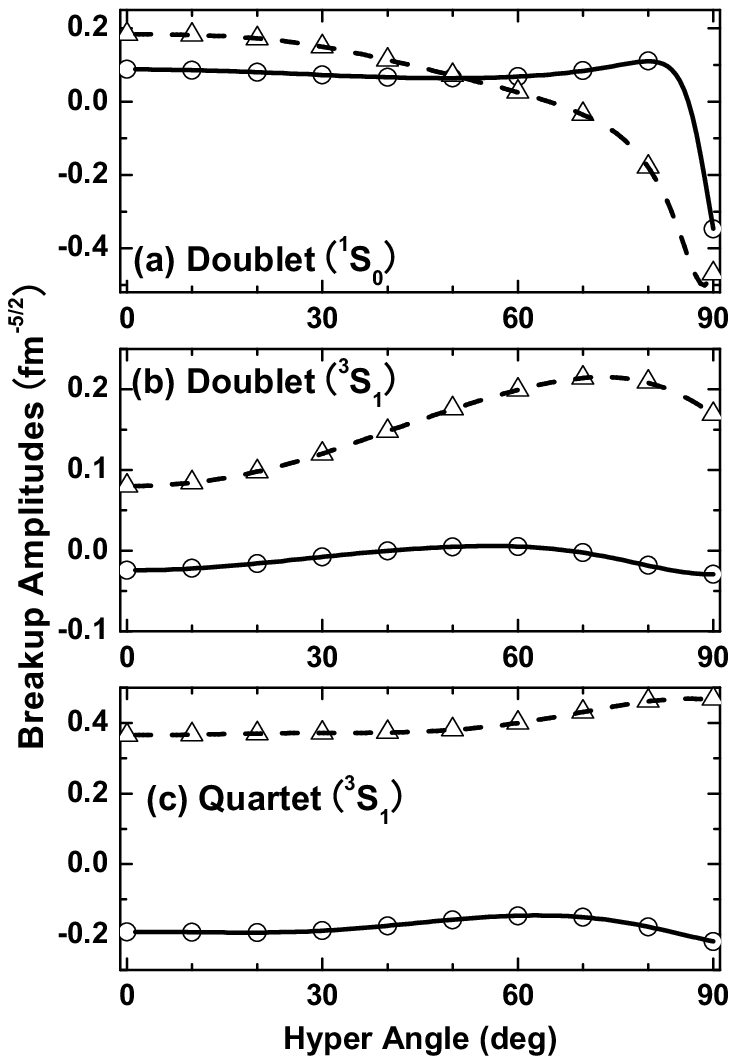}
\caption{
The real part (solid line) and the imaginary part (dashed line) of the breakup amplitudes at $E_{Lab}$=14.1 MeV in the present calculation.
Benchmark calculations in Ref.\ \protect\cite{Fr95} are plotted as circle and triangle points.
\label{fig:brk14mev}}
\end{minipage}
\hspace{\fill}
%
\begin{minipage}[t]{65mm}
\includegraphics[scale=0.8]{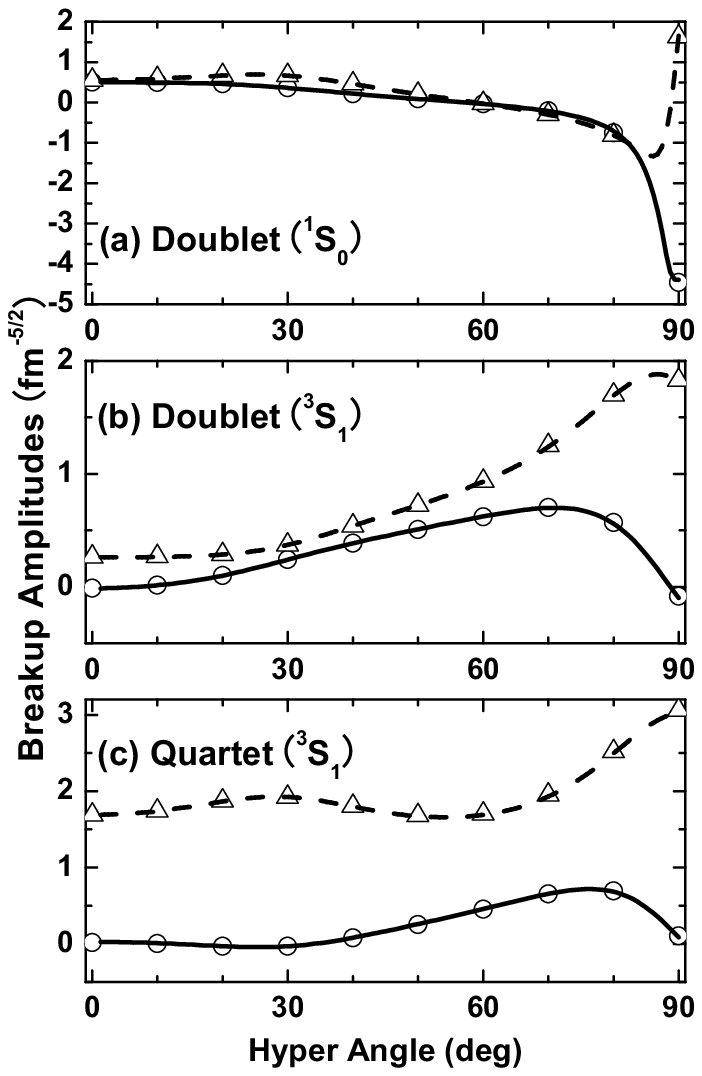}
\caption{
The real part (solid line) and the imaginary part (dashed line) of the breakup amplitudes at $E_{Lab}$=42 MeV in the present calculation.
Benchmark calculations in Ref.\ \protect\cite{Fr95} are plotted as circle and triangle points.
\label{fig:brk42mev}}
\end{minipage}
\end{figure}

The formalism for the operation of the Faddeev kernel described in the preceding sections is easily extended to more realistic cases, with spin degrees of freedom, with three-body forces, etc. 
Accommodating the formalism in the MCF algorithm \cite{Sa86,Is87}, 
we are able to solve the Faddeev integral equations in coordinate space.
To demonstrate the accuracy of our method, we performed calculations of the neutron-deuteron ($n$-$d$) scattering with the Malfliet-Tjon I-III potential, for which benchmark tests exist \cite{Fr90,Fr95}.
The comparison are made in Tables \ref{tab:nd-doub} and \ref{tab:nd-quar} and Figs. \ref{fig:brk14mev} and \ref{fig:brk42mev}.

%

In Tables \ref{tab:nd-doub} and \ref{tab:nd-quar}, where we tabulate results of the $s$-wave phase shift parameters for the $n$-$d$ doublet and quartet states at the incident energies $E_{Lab}$ of 4.0, 14.1, and 42.0 MeV, the calculations in the benchmark tests are denoted as Utrecht, J\"ulich/NM, Bochum, LA/Iowa, and Hosei(Q). 
(See Ref. \cite{Fr90} for further references of these methods.) 
In the calculations indicated as Utrecht and Bochum, coupled two-dimensional integral equations in momentum space are directly solved by Pad\'e approximant methods.
Integral kernels of their equations consist of free three-body Green's operator, two-body $t$-matrix, and permutations operators.
The two-body $t$-matrix possesses a pole due to the deuteron, whose effect is treated by a subtraction method.
The breakup effects appear as singularities in the three-body Green's function, see Ref. \cite{Gl96} for the details. 
Those indicated by J\"ulich/NM and Hosei(Q) use a separable expansion for two-body $t$-matrix to reduce the dimension of integral equations to one, and then solve the resulting equations taking into account of singularities in the kernels by technique of the contour deformation.
In the calculations denoted as LA/Iowa, the Faddeev differential equations in configuration space are solved with boundary conditions for the elastic and the breakup regions of the wave functions. 
We noted that the boundary condition for the elastic channel used there does not include the small oscillation behavior found in Fig. \ref{fig:elasticf}.

A breakup amplitudes defined in Ref. \cite{Fr95}, ${\cal A}(\Theta)$, is related with our amplitude $B_\alpha(\Theta)$ for $L=\ell=0$ as 
\begin{equation}
A(\Theta)= - e^{\frac{\pi}{4}\imath} \left( \frac{4}{3} \right)^{3/2}
  p_0 K_0^4 B_\alpha(\Theta).
\end{equation}
In Figs. \ref{fig:brk14mev} and \ref{fig:brk42mev}, the results of the breakup amplitude are compared for $E_{Lab}=$ 14.1 MeV and 42.0 MeV, respectively.
In the figures, our results for the real (imaginary) part are shown as the solid (dashed) curves, while those by Bochum and LA/Iowa groups \cite{Fr95}, which are almost equivalent, are denoted by circles (triangles).

All of our results agree with the benchmark calculations better than 1 \% level except about 2 \% discrepancy for the $\eta$ parameter in the quartet state at 42.0 MeV, which  demonstrates the present formalism is promising in solving the three-body scattering problem at energies above three-body breakup threshold.

\begin{table}
\beforetab
\begin{tabular}{lllllll}
\firsthline
$E_{Lab}$ (MeV)  & \multicolumn{2}{c}{4.0} & \multicolumn{2}{c}{14.1} & \multicolumn{2}{c}{42.0} \\
 & Re$(\delta)$ & $\eta$ & Re$(\delta)$ & $\eta$ & Re$(\delta)$ & $\eta$ \\
\midhline
Utrecht \protect\cite{Fr90} & 143.7 & 0.963 & 106.5 & 0.468 & 41.9 & 0.488 \\
J\"ulich/NM \protect\cite{Fr90} & 143.7 & 0.952 & 104.9 & 0.460 & 41.3 & 0.501 \\
Bochum \protect\cite{Fr90} & 143.7 & 0.964 & 105.5 & 0.467 & 41.3 & 0.504 \\
LA/Iowa \protect\cite{Fr90} & 143.7 & 0.964 & 105.4 & 0.463 & 41.2 & 0.501 \\
Hosei(Q) \protect\cite{Fr90} & 143.7 & 0.964 & 105.5 & 0.465 & 41.3 & 0.502 \\
This work &143.7  & 0.964 & 105.5 & 0.466 & 41.6 & 0.498\\
\lasthline
\end{tabular}
\aftertab
\captionaftertab[]{
Comparison of the benchmark calculations \protect\cite{Fr90} and the present calculations for neutron-deuteron spin-doublet phase shift parameters with the Malfliet-Tjon I-III potential.
\label{tab:nd-doub}
}
\end{table}

\begin{table}
\beforetab
\begin{tabular}{lllllll}
\firsthline
$E_{Lab}$ (MeV)  & \multicolumn{2}{c}{4.0} & \multicolumn{2}{c}{14.1} & \multicolumn{2}{c}{42.0} \\
 & Re$(\delta)$ & $\eta$ & Re$(\delta)$ & $\eta$ & Re$(\delta)$ & $\eta$ \\
\midhline
Utrecht \protect\cite{Fr90} & 102.1 & 1.000 & 68.8 & 0.978 & 38.4 & 0.898 \\
J\"ulich/NM \protect\cite{Fr90} & 101.1 & 1.000 & 68.5 & 0.986 & 37.2 & 0.907 \\
Bochum \protect\cite{Fr90} &  101.6 & 0.999 & 69.0 & 0.978 & 37.7 & 0.903 \\
LA/Iowa \protect\cite{Fr90} &  101.5 & 1.000 & 68.9 & 0.978 & 37.8 & 0.906\\
Hosei(Q) \protect\cite{Fr90} & 101.6 & 1.000 & 68.9 & 0.978 & 37.7 & 0.903 \\
This work & 101.6  & 1.000& 69.1 & 0.976 & 37.8 & 0.889\\
\lasthline
\end{tabular}
\aftertab
\captionaftertab[]{
Comparison of the benchmark calculations \protect\cite{Fr90} and the present calculations for neutron-deuteron spin-quartet phase shift parameters with the Malfliet-Tjon I-III potential.
\label{tab:nd-quar}}
\end{table}


\section{Summary}
\label{sec:summary}

We have presented a method to operate the Faddeev integral kernel in coordinate space at energies where three-body breakup reactions take place.
Effects of three-body breakup reactions appear as a long-range source contribution to the elastic component and to the breakup amplitudes at two-body sub-system having almost zero energy. 
Some numerical procedures are developed to treat these long-range behaviors.
With a model source function and a model potential, we have displayed some numerical examples to verify the accuracy of our method.

The procedure described in this paper can be used to solve the Faddeev equations in combination with an iterative algorithm to solve linear equations such as MCF. 
Solutions of the three-nucleon Faddeev equations are given for the Malfliet-Tjon I-III potential, and scattering phase shifts as well as the breakup amplitudes obtained from the solutions  give a good agreement with the benchmark solutions.
Results for three-nucleon systems with realistic nucleon-nucleon interactions and three-nucleon interactions will be presented elsewhere.

Since the integral kernel and hence the wave function in our formalism can be written as the sum of the elastic, three-body breakup, and closed channels, effects of each reaction mechanism can be easily drawn.
Our formalism thus can be extended to treat a three-body model of nuclear reactions including three-body breakup reactions in such a way that the theory resembles conventional theories of reactions.

\begin{acknowledge}
This research was supported by the Japan Society for the Promotion of Science, 
under Grants-in-Aid for Scientific Research No. 13640300.
The numerical calculations were supported, in part, by Research Center for Computing and Multimedia Studies, Hosei University, under Project No. lab0003.
\end{acknowledge}

\appendix

\section{Particle Exchange Operator $\hat{P}$}
\label{sec:OperatorP}

\setcounter{section}{1}

In this appendix, we summarize formulae to accomplish the particle exchange operator $\hat{P}$ in Eq. (\ref{eq:chi_xy}). 
$\chi_{\alpha}(x,y)$ in Eq. (\ref{eq:chi_xy}) is given as
\begin{equation}
\chi_{\alpha} (x,y) = \sum_{\alpha^\prime} \chi_{\alpha,\alpha^\prime}(x,y), 
\label{def_chi}
\end{equation}
where
\begin{eqnarray}
\chi_{\alpha,\alpha^\prime}(x,y) &=& 
   ({\cal Y}_\alpha  \vert \hat{P} \vert
   {\cal Y}_{\alpha^\prime} \xi_{\alpha^\prime}\rangle 
\nonumber \\
&=&
\sum_{a=0}^{L^\prime} \sum_{c=0}^{\ell^\prime} \sum_{b,d}
 \delta_{a+b, L^\prime}  \delta_{c+d, \ell^\prime} x^{a+c} y^{b+d}
 \sum_{\gamma,L_0}
K_\gamma^{\alpha^\prime}(x,y)
R_{(L \ell, L^\prime \ell^\prime) L_0}^{a c \gamma} 
\end{eqnarray}
with
\begin{eqnarray}
R_{(L \ell, L^\prime \ell^\prime) L_0}^{a c \gamma} &=& 
(-1)^{L_0+L^\prime-\ell^\prime+\gamma}
\hat{L} \hat{L^\prime} \hat{\ell} \hat{\ell^\prime} \hat{b} \hat{d}
\left(\begin{array}{c}
2L^\prime+1
\\
2a
\end{array} \right)^{1/2}
\left(\begin{array}{c}
2\ell^\prime+1
\\
2c
\end{array} \right)^{1/2}
\nonumber\\
&& \times 
\left( -\frac12 \right)^a  \left( 1 \right)^b
\left( -\frac34 \right)^c  \left( -\frac12 \right)^d
\sum_{e, f} (-)^{e+f} \hat{e} \hat{f}
\langle a c 0  0 \vert e 0 \rangle 
\langle b d 0  0 \vert f 0 \rangle 
\nonumber\\
&& \times 
\left\{
\begin{array}{ccc}
a & b & L^\prime \\
c & d & \ell^\prime \\
e & f & L_0
\end{array}
\right\}
\langle L e 0 0 \vert\gamma0\rangle 
\langle \ell f 0 0 \vert \gamma 0 \rangle
\left\{
\begin{array}{ccc}
e    & \gamma & L \\
\ell & L_0    & f
\end{array}
\right\}
\end{eqnarray}
and
\begin{equation}
K_\gamma^{\alpha^\prime}(x,y) = \int_{-1}^1 du
 \frac{\xi_{\alpha^\prime}(x^{\prime\prime}, y^{\prime\prime})}
    {(x^{\prime\prime})^{L^\prime} (y^{\prime\prime})^{\ell^\prime} }
    P_\gamma(u).
\label{eq:K_gamma}
\end{equation}
Here, $\hat{n}$ denotes $\sqrt{2n+1}$; $P_\gamma(u)$ is the Legendre polynomial; $x^{\prime\prime}$ and
 $y^{\prime\prime}$ are
\begin{equation}
\left\{
\begin{array}{lll}
x^{\prime\prime} &=& \sqrt{ \frac14 x^2 - x y u + y^2 }
\\
y^{\prime\prime} &=& \sqrt{ \frac9{16} x^2 + \frac34 x y u + \frac14 y^2 }.
\end{array}
\right.
\label{eq:xy-prime}
\end{equation}

\section{Green's operator}
\label{sec:Green-x}

In this appendix, we first review two-body Green's operators and describe how to calculate Eq.\ (\ref{eq:eta-Gomega}).

We define Green's operators for the outgoing $(+)$ and the incoming $(-)$ boundary conditions with and without a potential as
\begin{equation}
G_{L}^{(\pm)} =  \frac1{E_q \pm \imath \varepsilon - T_{L}(x) - V(x)} 
\label{eq:Green_pm}
\end{equation}
\begin{equation}
G_{0,L}^{(\pm)} =  \frac1{E_q \pm \imath \varepsilon - T_L(x)}.
\end{equation}
These satisfy resolvent relations
\begin{equation}
G_{L}^{(\pm)} 
  = G_{0,L}^{(\pm)} + G_{L}^{(\pm)} V G_{0, L}^{(\pm)} 
  = G_{0,L}^{(\pm)} + G_{0,L}^{(\pm)} V G_{L}^{(\pm)}.
\label{eq:resolvent}
\end{equation}

Two-body scattering wave functions corresponding to the outgoing and the incoming boundary conditions $\vert \psi_L^{(\pm)}\rangle$ satisfy the Lippmann-Schwinger equations
\begin{equation}
\vert \psi_L^{(\pm)}\rangle = \vert j_L \rangle
   + G_{0,L}^{(\pm)} V \vert \psi_L^{(\pm)}\rangle,
\end{equation}
whose formal solutions are written as
\begin{equation}
\vert \psi_L^{(\pm)}\rangle = \vert j_L \rangle
   + G_{L}^{(\pm)} V \vert j_L \rangle.
\label{eq:formal_sol}
\end{equation}

Although the Green's operators and the wave functions above are complex values, we do not necessarily have to handle complex values when the potential $V(x)$ is real. 
For this we define the principal values of the two-body Green's operators ${\cal P}G_{L}$ and ${\cal P}G_{0,L}$
\begin{equation}
{\cal P}G_{L}   = {\cal P} \frac1{E_q - T_L(x) - V(x)}
\end{equation}
\begin{equation}
{\cal P}G_{0,L}   = {\cal P} \frac1{E_q - T_L(x)}.
\end{equation}
As is $G_{0,L}^{(\pm)}$, an analytical form of ${\cal P}G_{0,L}^{(\pm)}$ is known and these operators are related as
\begin{equation}
G_{0,L}^{(\pm)} =  {\cal P}G_{0,L} \mp \imath q \frac{m}{\hbar^2} \vert j_L \rangle \langle j_L \vert,
\label{eq:G0PG0}
\end{equation}

A scattering wave function corresponding to ${\cal P}G_{0,L}$, namely standing wave solution $\vert\hat{\psi}_L \rangle$ satisfies
\begin{equation}
\vert \hat{\psi}_{L} \rangle = \vert j_L \rangle 
 + {\cal P}G_{0,L} V \vert \hat{\psi}_{L} \rangle,
\label{eq:LS-standing}
\end{equation}
and a formal solution of this is given as 
\begin{equation}
\vert \hat{\psi}_L \rangle 
  = \vert j_L \rangle  + {\cal P}G_{L} V \vert j_L \rangle.
\label{eq:formal_sol_0}
\end{equation}

From the standing wave solution, the outgoing and the incoming solutions are obtained as
\begin{equation}
\vert \psi^{(\pm)}_{L} \rangle = 
  \frac1{1 \mp \imath {\cal K}_L } \vert \hat{\psi}_{L} \rangle,
\end{equation}
where ${\cal K}_L$ is the scattering $K$-matrix defined by
\begin{equation}
{\cal K}_{L} 
  = - q \frac{m}{\hbar^2} \langle j_L \vert V \vert \hat{\psi}_{L} \rangle,
\end{equation}
which becomes $\tan\delta$ with a phase shift parameter $\delta$.
Using the relations above, one obtains a relation between $G_{L}^{(\pm)}$ and ${\cal P}G_{L} $ as
\begin{equation}
G_{L}^{(\pm)}  = {\cal P}G_{L} 
 \mp \imath q \frac{m}{\hbar^2} \vert \hat{\psi}_{L} \rangle 
   \frac1{1 \mp \imath {\cal K}_{L}} \langle \hat{\psi}_{L} \vert,
\label{eq:GcalG}
\end{equation}
which reduces to Eq.\ (\ref{eq:G0PG0}) if $V(x)$ was 0, leading to $\hat{\psi}_{L}(x) = j_L(qx)$ and ${\cal K}_{L} = 0$.

Next, we discuss about asymptotic form of the Green's functions.
The asymptotic forms of $G_{0,L}^{(\pm)}$ and ${\cal P}G_{0,L}$ are obtained from their analytical forms  as 
\begin{equation}
G_{0,L}^{(\pm)} 
  \to   - q \frac{m}{\hbar^2}  \vert h_L^{(\pm)}\rangle \langle j_L \vert,
\label{eq:G0_asym}
\end{equation}
\begin{equation}
{\cal P}G_{0,L} \to  q \frac{m}{\hbar^2} \vert n_L \rangle \langle j_L \vert.
\label{eq:PG0asym}
\end{equation}
These equations and the resolvent equations together with the formal solutions Eqs. (\ref{eq:formal_sol}) and (\ref{eq:formal_sol_0}) lead to
\begin{equation}
G_{L}^{(\pm)} \to 
 - q \frac{m}{\hbar^2} \vert h_L^{(\pm)}\rangle \langle \psi_L^{\mp} \vert, 
\end{equation}
\begin{equation}
{\cal P}G_{L}  \to  
  q \frac{m}{\hbar^2} \vert n_L \rangle \langle\hat{\psi}_{L} \vert.
\label{eq:calG_asym}
\end{equation}

Finally, we describe how to calculate Eq.\ (\ref{eq:eta-Gomega}), which we write simply as
\begin{equation}
 \eta(x) = \langle x \vert G_{L}^{(+)} \vert \hat{\omega} \rangle.
\label{eq:etaGomega}
\end{equation}

Using Eq.\ (\ref{eq:GcalG}), one can write $\eta(x)$ as
\begin{equation}
\eta(x) = \bar{\eta}(x) 
- \imath q \frac{m}{\hbar^2} \hat{\psi}_{L}(x)
   \frac1{1 - \imath {\cal K}_{L}} \langle \hat{\psi}_{L} \vert \hat{\omega} \rangle,
\end{equation}
where a new function $\bar{\eta}(x)$ is defined by
\begin{equation}
 \bar{\eta}(x)  = \langle x \vert {\cal P}G_{L} \vert \hat{\omega} \rangle.
\end{equation}
From Eq.\ (\ref{eq:calG_asym}), the asymptotic form of $\bar{\eta}(x)$ can be written as
\begin{equation}
\bar{\eta}(x)  \mathop{\to}_{x \to \infty}  
  q \frac{m}{\hbar^2} n_L (qx) \langle\hat{\psi}_{L} \vert \omega \rangle
\end{equation}

In actual calculation, the function $\bar{\eta}(x)$ is obtained by solving the ordinary differential equation
\begin{equation}
\left[ E_q  - T_L(x) - V(x) \right] \bar{\eta}(x)  = \hat{\omega}(x)
\end{equation}
with the boundary condition
\begin{equation}
\bar{\eta}(x)  \mathop{\propto}_{x \to \infty}   n_L (qx).
\end{equation}

These relations give the asymptotic form of $\eta(x)$ as
\begin{equation}
\eta(x)  \mathop{\to}_{x \to \infty} 
 h_L^{(+)}(qx) \frac1{1- \imath {\cal K}_{L}} \left( -q \frac{m}{\hbar^2} \right)
 \langle \hat{\psi}_{L} \vert \hat{\omega} \rangle.
\end{equation}

\section{Mesh points for $x$ and $y$ variables}
\label{sec:meshes}

Crucial procedures in our numerical calculations are to solve the differential equations Eq.\ (\ref{eq:SchFad}) and the Fourier-Bessel transformation Eq.\ (\ref{eq:chi_xp}), which are related to $x$- and $y$-mesh points, respectively.
In this appendix, we give some remarks on these mesh points.

Both mesh points are taken in uneven distances so as to be shorter near the origin to take into account of short range nuclear potentials.

Uneven mesh points, for $x$-mesh, e.g., are created with the same functional form as the one used in Ref.\ \cite{Sa81} 
\begin{equation}
t(x) = \frac{c (x + t_0) x}{x + s_0}
\label{eq:t-of-x}
\end{equation}
or inversely
\begin{equation}
x(t) = \frac{-(ct_0-t)-\sqrt{(ct_0-t)^2+4cs_0t}}{2c},
\end{equation}
with equidistant $t$-mesh points.
The parameters of  Eq.\ (\ref{eq:t-of-x}), $c$, $t_0$, and $s_0$, are determined from the following conditions:
\begin{enumerate}
\item
The $x$-mesh size near the origin : $\Delta x_0$
\item
The $x$-mesh size at large distance (the infinity) : $\Delta x_\infty$
\item
A value of $x$-mesh points, $x_m$, and the number of the mesh points for $0 < x \le x_m$ : $N_x$
\end{enumerate}

We can choose the distance of $t$-mesh points, $\Delta t$,  as arbitrary, and thus 
\begin{equation}
t_m  = N_x \Delta t = t(x_m)
\end{equation}

From Eq.\ (\ref{eq:t-of-x}), 
\begin{eqnarray}
\left. \frac{dt}{dx}\right\vert_{x=0} = \frac{c t_0}{s_0}
\\
\left. \frac{dt}{dx}\right\vert_{x=\infty} = c
\end{eqnarray}

Then,
\begin{eqnarray}
\Delta x_0 &=& \frac{s_0}{c t_0} \Delta t
\\
\Delta x_\infty &=&\frac1c\Delta t 
\end{eqnarray}

Using the values of $N_x$, $x_m$, $\Delta x_0$, and $\Delta x_\infty$ as an input, we rewrite the above conditions as
\begin{eqnarray}
c &=& \frac{\Delta t}{\Delta x_\infty}
\\
s_0 &=& \frac{N_x-\frac{x_m}{\Delta x_\infty}}{\frac{x_m}{\Delta x_0}-N_x}x_m
\\
t_0 &=& \frac{\Delta x_\infty}{\Delta x_0} s_0
\end{eqnarray}

In the present calculations, we set $\Delta t=1$  (fm) both for $x$- and $y$-mesh points;
 $N_x=60 (100)$, $x_m = 10 (80)$ fm, $\Delta x_0=0.025 (0.033)$, $\Delta x_\infty=0.3 (1.25)$ for $x$- ($y-$) mesh points.


\end{document}